\documentclass[epj,referee]{svjour}
\usepackage{psfig}
\newcommand{\beq}{\begin{equation}}
\newcommand{\eeq}[1]{\label{#1} \end{equation}}
\newcommand{\insertplot}[1]{\centerline{\psfig{figure={#1},width=17.9cm}}}
\newcommand{\insertplots}[1]{\centerline{\psfig{figure={#1},width=8.9cm}}}
\newcommand{\insertplotss}[1]{\centerline{\psfig{figure={#1},width=7cm}}}

\begin{document}
\title{Resonance-reggeon and parton-hadron duality
 in strong interactions}
\author{L. Jenkovszky\inst{1} \and V.K. Magas\inst{1,2}
\and E. Predazzi\inst{3}
}                     
%
%
\institute{
Bogolyubov Institute for Theoretical Physics, Academy of Sciences of Ukraine\\
Metrologicheskaya st. 14b, 01143 Kiev, Ukraine
\and
Center for Physics of Fundamental Interactions, Instituto Superior Tecnico\\
Av. Rovisco Pais, 1049-001 Lisbon, Portugal
\and
Torino University and INFN\\
via P. Giuria, 1, Torino, 10125 Italy
}
\date{Received: date / Revised version: date}
%
\abstract{
By using the concept of duality between direct channel
resonances and Regge exchanges we relate the small- and large-$x$
behavior of the structure functions. We show that even a single
resonance exhibits Bjorken scaling at large $Q^2$.
} 
\maketitle
\section{Introduction}
\label{intro}
Inspired by recent experimental measurements of the nucleon
structure functions at the JLab (CEBAF) \cite {Niculescu}, we
suggest a unified "two-dimensionally dual" picture of the strong
interaction \cite{JMP,FJM,JM} connecting low- and high energies (Veneziano, or
resonance-reggeon duality \cite{Veneziano}) with low- and high
virtualities ($Q^2$)
(Bloom-Gilman, or hadron-parton duality \cite{BG}).
The basic idea of the unification is the use of 
$Q^2$-dependent dual amplitudes,
employing nonlinear complex Regge trajectories providing an
imaginary part of the scattering amplitude related to the total
cross section and structure functions,  thus saturating the duality by a
finite number of resonances lying on the (limited) real part of
the Regge trajectories.

\begin{figure*}[htb]
        \insertplot{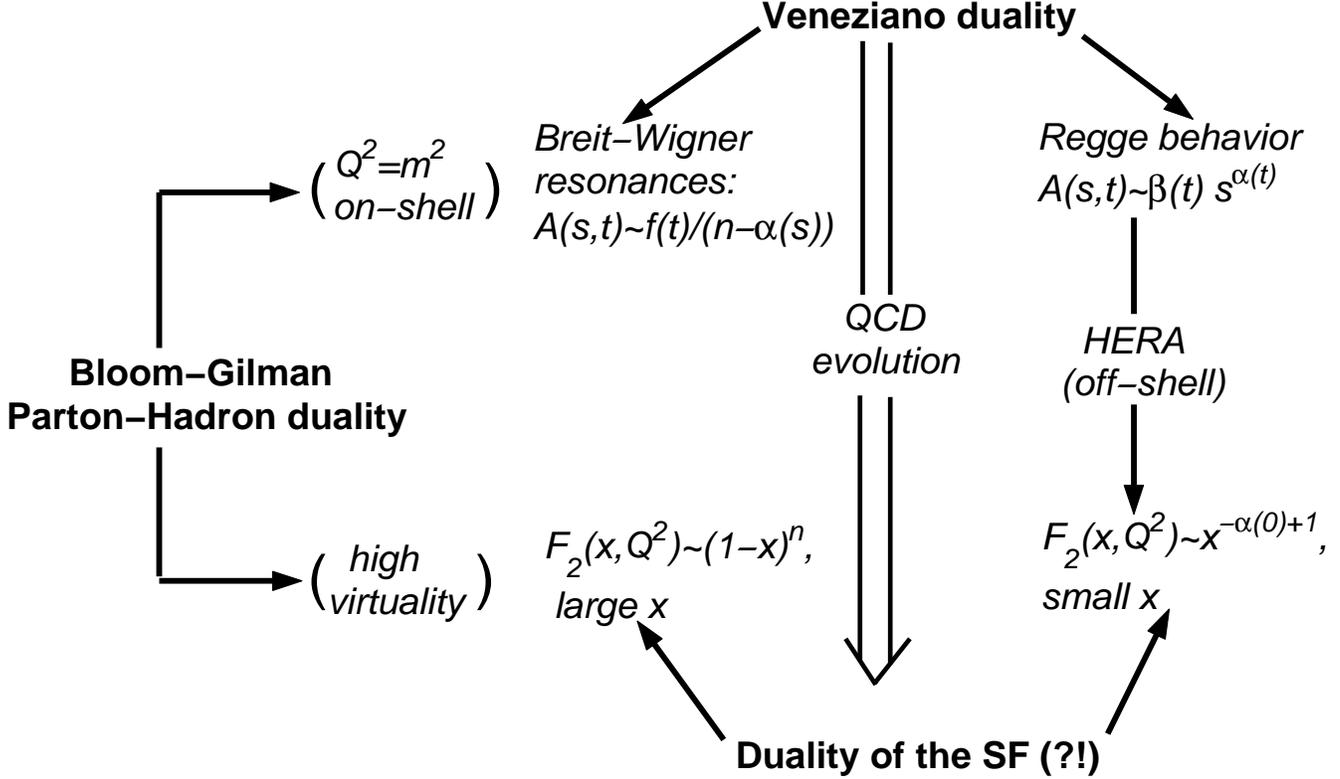}
\caption{Veneziano, or resonance-reggeon,
duality \cite{Veneziano} and
Bloom-Gilman, or  hadron-parton, duality \cite{BG}
 in strong interactions}
\label{diag}
\end{figure*}

The resulting object, a deeply virtual
scattering amplitude, $A(s,t,Q^2)$, is a function of three
variables, reducing to the nuclear structure function (SF), when
$t=0$, and  on-shell hadronic scattering amplitude for
$Q^2=m^2$. It closes the circle in Fig \ref{diag}. We use this amplitude to describe the
background as well as the resonance component \cite{FH}.

The $Q^2-$ dependence of the residue functions here will be
chosen in such a way as to provide us with Bjorken scaling at small
$x$ (large $s$). The resulting amplitude (structure function) is
applicable in the whole kinematical range,  resonance
region included . We call this unification "two dimensional duality" - one
in $s$, the other one in $Q^2$,

At the early days of duality, off-mass continuation was attempted
\cite{RR} by means of multi-leg (e.g. 6-point) dual amplitudes
with "extra" lines taken at their poles. Without going into
details, here we only mention that scaling in this approach can be
achieved \cite{Schierholz} with nonlinear trajectories only, e.g.
trajectories  with logarithmic or constant asymptotics.

\section {Notation and conventions}
\begin{figure}[htb]
        \insertplotss{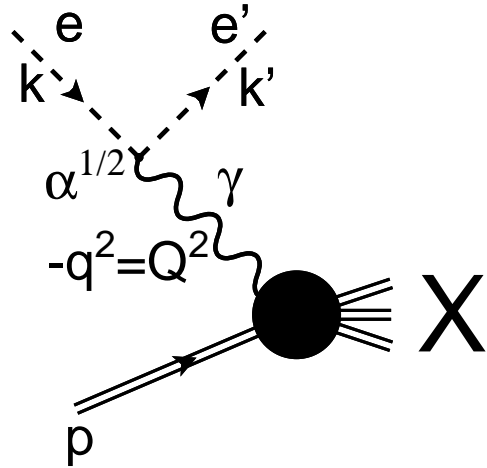}
\caption{Kinematics of deep inelastic scattering}
\label{d1}
\end{figure}

We use standard notation for the cross section and structure
function (see Fig. \ref{d1}):
 \beq
 \sigma^{\gamma^* p}={4\pi^2\alpha(1+4m^2x^2/Q^2
)\over{Q^2(1-x)}}{F_2(x,Q^2)\over{1+R(x,Q^2)}}, \eeq{eq1} where
$\alpha$ is the fine structure constant, $Q^2$ is the
four-momentum transfer squared (with  minus sign) or the momentum carried 
by the virtual photon, $x$ is the Bjorken variable and $s$ is the
centre-of-mass energy squared  of the $\gamma^*p$ system obeying the
relation as follows:
\beq s=Q^2(1-x)/x+m_p^2, \eeq{eq2} where $m_p$ is the
proton mass and $R(x,Q^2)=\sigma_L(x,Q^2)/\sigma_T(x,Q^2)$. For
the sake of simplicity we set $R=0$, that is a reasonable
approximation.

\begin{figure*}[htb]
        \insertplot{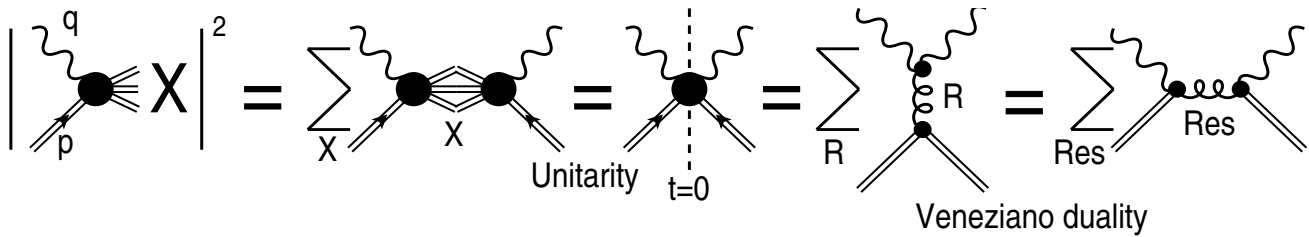}
\caption{According to the Veneziano (or resonance-reggeon) duality a proper sum
of either t-channel or s-channel resonance exchanges accounts for
the whole amplitude}
\label{d2}
\end{figure*}

We use the norm where \beq \sigma_T^{\gamma^*}(s,t,Q^2)=Im\
A(s,t,Q^2). \eeq{eq3} According to the two-component duality
picture \cite{FH}, both the scattering amplitude $A$ and structure
function $F_2$ are the sums of the diffractive and non-diffractive
terms. At high energies both terms are of the Regge type. For $\gamma^*
p$ scattering the positive-signature exchanges are allowed only. The
dominant ones are the Pomeron and  $f$ Reggeon, respectively.
The relevant scattering amplitude is as follows (here $t=0$):
\beq
A_i(s,Q^2)=i\beta_k(Q^2)\Bigl(-i{s\over{s_i}}\Bigr)^{\alpha_k(0)-1},
\eeq{eq4} 
where $\alpha$ and $\beta$ are Regge trajectories and
residues and $k$ stands either for the Pomeron or the Reggeon. As usual, the
residue is chosen  to satisfy approximate Bjorken
scaling for the structure function \cite{BGP,K}. 
Assuming the Reggeon (or Pomeron) exchange to be a simple pole, 
the residue function obeys the  factorization property: it is a
product of two vertices - the $\gamma\gamma R(P)$ and $NNR(P)$,
where $N$ stands for the nucleon (see Fig. \ref{d2}).

At low energies the scattering amplitude is dominated by the
contribution of the near-by resonances. In the vicinity of the resonance,
$Res$, the amplitude can be also written in a factorized form -- as a
product of probabilities that two particles, $\gamma$ and $p$,
form a resonance with the mass squared $s_{Res}$ and total width $\Gamma$:
\beq
A(s,Q^2)=\sum_{spin}{A_{fi}(Q^2)A^*_{if}(Q^2)\over{s_{Res}-s-i\Gamma}},
\eeq{eq5} 
where $A_{fi}$ are the inelastic form
factors.

\section{Nucleon resonances in inelastic electron-nucleon
scattering}
About thirty years ago Bloom and Gilman \cite{BG}
observed that the prominent resonances in  inelastic
electron-proton scattering do not disappear with increasing $Q^2$
with respect to the "background" but instead fall at roughly the
same rate as any background. Furthermore, the smooth scaling limit
proved to be an accurate average over resonance bumps seen at
lower $Q^2$ and $s$.

Since then, the phenomenon was studied in a number of papers
\cite{carlson,Carl} and recently has been confirmed experimentally
\cite{Niculescu}. These studies were aimed
mainly to answer the
questions: in which way a  limited number of resonances can reproduce
the smooth scaling behaviour? The main
theoretical tools in these studies were finite energy sum rules
and perturbative QCD calculations, whenever applicable. Our aim
instead is the construction of an explicit dual model combining
direct channel resonances, Regge behaviour typical for hadrons and
scaling behaviour typical for the partonic picture.

The  existence of resonances in the structure function at large
$x$ close to $x=1$ is not surprising by itself: as it follows from
(\ref{eq1}) and (\ref{eq2}) they are the same as in $\gamma^*p$
total cross section, but in a different coordinate system. The important
question is whether and, if so, how a small number  of resonances (or even a
single one) can reproduce the smooth Bjorken scaling behaviour,
known to be an asymptotic property which is typical for multiparticle
processes.

The possibility that a limited (small) number of resonances can
build up the smooth Regge behaviour was demonstrated by means of finite
energy sum rules \cite{DHS}. Later it was confused by the presence of 
an infinite number of narrow resonances in the Veneziano model 
\cite{Veneziano}, which made its phenomenological application
difficult, if not impossible. Similar to the case of the
resonance-reggeon duality \cite{DHS}, the hadron-parton duality was
established \cite{BG} by means of the finite-energy sum rules,
but it was not realized explicitly like the Veneziano
model (or its further modifications).

Actually, the early onset of Bjorken scaling, which is called "early, or
precaution scaling", was observed with the first measurements of
deep inelastic scattering at SLAC, where it was noticed that a
more rapid approach to scaling can be achieved with the Bloom-Gilman (BG)
variable \cite{BG} $x'=x/(1+m^2x^2/Q^2)$ instead of $x$ (or
$\omega=1/x$). More recently the following generalization of the
BG variable, such as
\beq
\xi={2x\over{1+\sqrt{1+{4m^2x^2\over Q^2}}}}\ ,
\eeq{nach}
was suggested by O.Nachtmann \cite{Nachtmann}. We use the standard
Bjorken variable $x$, however our results can be easily rewritten in terms
of the above-mentioned modified variables.

First attempts to combine resonance (Regge) behaviour with
Bjorken scaling were made \cite{DG,BEG,EM} at low energies (large
$x$), with the emphasis on the right choice of the
$Q^2$-dependence, such as to satisfy the needed behaviour of 
form factors, vector meson dominance (VMD) with the requirement of 
Bjorken scaling. (N.B.: the
validity (or failure) of the (generalized) VMD
is still disputed). Similar attempts in the high-energy (low $x$)
region became popular recently, with the advent of the HERA data.
They are presented in Sec. 5.

A consistent treatment of the problem requires the account for the
spin dependence. For the sake of simplicity we ignore it in this paper (see 
e.g.  \cite{Carl}).

\section{Factorization and dual properties (bootstrap)
of the vertices}
Since the purpose of the present paper is the construction of a
unified model realizing duality both in the $s$ and $t$ channels,
we first attempt to identify its fragments, namely, the vertices
(to be interpreted later on as $Q^2$-dependent form factors).

Let us remind that the residue functions are completely arbitrary
in the Regge pole model, but they are constrained in the dual
model. We show this by using the low-energy (resonances) and 
high-energy (Regge) decomposition in the simple Veneziano model
\cite{Veneziano} $$V(s,t)=\int_0^1 dz
z^{-\alpha(s)}(1-z)^{-\alpha(t)}= $$ \beq
B(1-\alpha(s),1-\alpha(t)) =
{\Gamma(1-\alpha(s))\Gamma(1-\alpha(t))
\over{\Gamma(2-\alpha(s)-\alpha(t))}}. \eeq{eq11} 
Furthermore,
\beq
V(s,t)=\sum_{n=1}^{\infty}{1\over{n-\alpha(s)}}{\Gamma(n+\alpha(t)+1)
\over {n!\ \Gamma(\alpha(t)+1)}}. \eeq{eq12}

By the Stirling formula
$$
 V(s,t)\Bigg|_{|\alpha(s)|\rightarrow\infty}\rightarrow
[-\alpha(s)]^ {\alpha(t)-1} \Gamma\bigl(1-\alpha(t)\bigr)
$$
\beq
 \biggl[\sum_{n=0}^N{a_n(0) \over
{[\alpha(s)]^n}}+0\biggl({1\over{[\alpha(s)]^{N+1}}}\biggr)
\biggr]\ , \eeq{eq13} 
and, since for small $|t|$ the $\Gamma$ function
varies slowly compared with the exponential one, the Regge asymptotic
behaviour is \beq V(s,t)\sim (\alpha' s)^{\alpha(t)}, \eeq{eq14}
where $\beta(t)=(\alpha')^{\alpha(t)}$ is the Regge residue.

Actually, one has to identify a single (and hence economic!) Regge
exchange amplitude with a sum of direct-channel poles. Such an
identification is not practical for infinite number of poles
(e.g. the Veneziano amplitude) but, as we show below, is feasible
if their number is finite (small). To anticipate the forthcoming
discussion, we feed the $Q^2$-dependence in the Regge
residue at high energies (small $x$) and use the dual amplitude
(with  finite number of resonances!) for the whole kinematical
region, including that of resonances. Relating the amplitude to
the SF, we set $t=0$.

To remedy the problems of the infinite number of narrow resonance,
nonunitarity and lack of the imaginary part, we use a
generalization of the Veneziano model free from the
above-mentioned difficulties.

\section{Dual amplitude with Mandelstam analyticity}
The invariant dual on-shell scattering
amplitude with Mandelstam analyticity (DAMA)
applicable  both to the diffractive and non-diffractive components
reads \cite{D}: \beq D(s,t)=\int_0^1 {dz \biggl({z \over g}
\biggr)^{-\alpha(s')-1} \biggl({1-z \over
g}\biggr)^{-\alpha(t')-1}}, \eeq{eq21} where $s'=s(1-z), \ \
t'=tz, \ \ g$ is a parameter, $g>1$, and $s, \ \ t$ are the
Mandelstam variables.

For $s\rightarrow\infty$ and fixed $t$ it has the following Regge
asymptotic behaviour \beq
D(s,t)\approx\sqrt{{2\pi\over{\alpha_t(0)}}}g^{1+a+ib}\Biggl({s\alpha'(0)g\ln
g\over{\alpha_t(0)}}\Biggl)^{\alpha_t(0)-1}, \eeq{eq22} where
$a=Re\ \alpha\Bigl({\alpha_t(0)\over{\alpha'(0)\ln g}}\Bigr)$ and
$b=Im\ \alpha\Bigl({\alpha_t(0) \over{\alpha'(0)\ln g}}\Bigr)$.

The pole structure of DAMA is similar to that of the Veneziano
model except that multiple poles may appear at daughter levels.
The presence of these multipoles does not contradict the
theoretical postulates.  On the other hand, they can be removed
without any harm to the dual model by means the so-called Van der
Corput neutralizer. The procedure  \cite{D} is to multiply the
integrand of (\ref{eq21}) by a function $\phi(x)$ with the
properties:
$$ \phi(0)=0,\ \ \ \phi(1)=1,\ \ \ \phi^n(1)=0,\ \ n=1,2,3,... $$
The function $ \phi(x)=1-\exp\Biggl({-{x\over{1-x}}}\biggr), $ for
example, satisfies the above conditions and results \cite{D} in a
standard, "Veneziano-like" pole structure: \beq
D(s,t)=\sum_ng^{n+\alpha_t(0)}{C_n\over{n-\alpha(s)}}, \eeq{eq23}
where \beq
C_n={\alpha_t(0)\Bigl(\alpha_t(0)+1\Bigr)...\Bigl(\alpha_t(0)+n+1\Bigr)\over{n!}}.
\eeq{eq24}

The pole term (\ref{eq23}) is a generalization of the Breit-Wigner
formula (\ref{eq5}), comprising a whole sequence of resonances
lying on a complex trajectory $\alpha(s)$. Such a "reggeized"
Breit-Wigner formula has little practical use in the case of
linear trajectories, resulting in an infinite sequence of poles,
but it becomes a powerful tool if complex trajectories with a
limited real part and hence a restricted number of resonances are
used. Moreover, it appears that a small number of resonances is sufficient
to saturate the direct channel.

Contrary to the Veneziano model, DAMA (\ref{eq21}) not only
allows but rather requires the use of nonlinear complex
trajectories providing, in particular, for the imaginary part of
the amplitude, resonance widths and resulting in a finite number
of those. More specifically, the asymptotic rise of the
trajectories in DAMA is limited by the condition (in accordance
with an important upper bound derived earlier \cite{DP}): \beq
|{\alpha(s)\over{\sqrt s\ln s}}|\leq const, \ \
s\rightarrow\infty. \eeq{lim} Actually, this upper bound can be
even lowered up to a logarithm by requiring wide angle scaling
behaviour for the amplitude.

The models of Regge trajectories combining the correct threshold and
asymptotic behaviours have been widely discussed in the literature
(see e.g. \cite{FJMPP} for a recent treatment of this problem). A
particularly simple model is based on a sum of square roots
$$\alpha(s)=\alpha_0+\sum_i\gamma_i \sqrt{s_i-s}, $$
 where the
lightest threshold (made of two pions or a pion and a nucleon) is
important for the imaginary part, while the heaviest threshold
limits the rise of the real part, where resonances terminate.

Dual amplitudes with Mandelstam 
analyticity with  trajectories specified above are equally applicable
to both: the diffractive and non-diffractive components of the
amplitude, the difference being qualitative rather than
quantitative. The utilization of a trajectory with a single
threshold, \beq
\alpha_E(s)=\alpha_E(0)+\alpha_{1E}(\sqrt{s_E}-\sqrt{s_E-s})
\eeq{eq32} prevents the production of resonances on the the
physical sheet, although they are present on the nonphysical
sheet, sustaining duality (i.e. their sum produces Regge
asymptotic behaviour). This nontrivial property of DAMA makes it
particularly attractive in describing the smooth background (dual
to the Pomeron exchange) (see \cite{D}). The threshold value, slope and 
intercept of this exotic trajectory are free parameters.

For the resonance component a finite sum  in (\ref{eq23}) is
adequate, but we use a simple model with the lowest
threshold included explicitly and the higher ones approximated by
a linear term: \beq
\alpha_R(s)=\alpha_R(0)+\alpha's+\alpha_{1R}(\sqrt{s_0}-\sqrt{s_0-s}),
\eeq{eq33} where $s_0$ is the lowest threshold --
$s_0=(m_{\pi}+m_p)^2$ in our case -- while the remaining 3
parameters are adjusted to the known properties of 
relevant trajectories ($N^*$ and $\Delta$ isobar in our case).
 The termination of resonances, which are  provided in DAMA by the limited
real part, are effectively taken into account here by a cutoff
in the summation of (\ref{eq23}).

Finally, we note that a minimal model for the scattering amplitude
is a sum \beq A(s,t,u)=c(D(s,t)+D(u,t)), \eeq{eq34} providing
the correct signature at high-energy limit, $c$ is a normalization
factor. We disregard the symmetry (spin and isospin) properties of
the problem, concentrating on its dynamics. In the limit
$s\rightarrow\infty,\ \ t=0$ we have $u=-s$ and therefore \beq
A(s,0,-s)|_{s\rightarrow\infty}=c\ D(s,0)(1+(-1)^{\alpha_t(0)-1}),
\eeq{eq35a} where $D(s,t)$ is given by eq. (\ref{eq22}). For the
total cross section in this limit we obtain:
$$
\sigma_T^{\gamma^*}=Im\ A=Cg^{\alpha_t(0)+a} \left(s \alpha'(0)
\ln g \right)^{\alpha_t(0)-1}\cdot$$ \beq \cdot\left(\sin
(\alpha_t(0)-1)\pi \cos (b\ln g)+\right.\eeq{eq35}
$$\left.+(1+\cos (\alpha_t(0)-1)\pi)\sin (b\ln g) \right)\ ,$$
where $C$ is a constant independent of $s,\ g$ and $\alpha'(0)$.

\section {$Q^2-$ dependence}
Our main idea is to introduce the  $Q^2$-dependence in the
dual model by matching its Regge asymptotic behaviour and pole
structure to standard forms known from the literature. The point
is that the correct identification of this $Q^2$-dependence in a
single asymptotic limit of the dual amplitude will extend it to
the rest of the kinematical regions. We have two ways to do so, that is,\\
A) to combine  Regge behaviour and Bjorken scaling limits of the
structure functions (or $Q^2$-dependent $\gamma^*p$ cross
sections), or\\
B) to introduce properly $Q^2$ dependence in the resonance region.\\
They should match to each other, if the procedure is correct, and the dual
amplitude should take care of any further inter- or
extrapolation.

\begin{figure}[htb]
        \insertplots{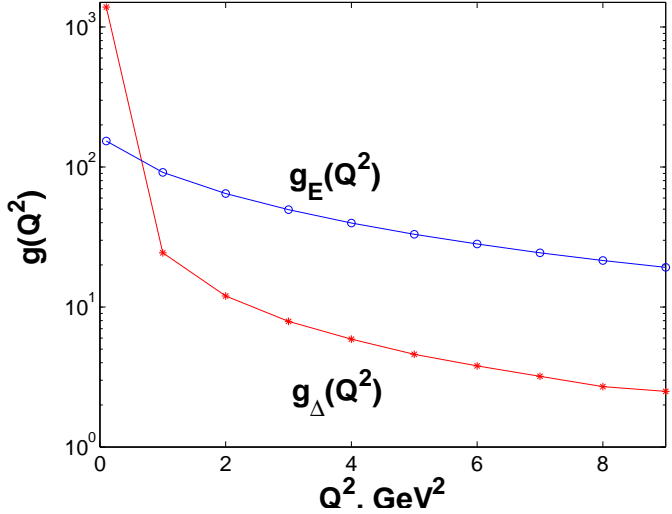}
\caption{$g(Q^2)$ - the solution of the transcendent equation
(\ref{eq45}) - for $\Delta$ and exotic
trajectories.}
\label{fig1}
\end{figure}

It is obvious from eq. (\ref{eq4}) that asymptotic Regge and
scaling behaviour require the residue to fall like
$\sim(Q^2)^{-\alpha_i(0)+1}$. Actually, it could be more involved
if we require the correct $Q^2\rightarrow 0$ limit to be respected
and the observed scaling violation (the "HERA effect") to be
included. Various models to cope with the above requirements have
been suggested \cite{BGP,K,JMP99}. At HERA, especially at large
$Q^2$, scaling is so badly violated that it may not be explicit
anymore.

In combining Regge asymptotic behaviour with (approximate) Bjorken
scaling, one can proceed basically in the following way -- to keep
explicitly a scaling factor $x^{\Delta}$ (to be broken by some
$Q^2$-dependence "properly" taken into account) \cite{K}

\beq F_2(x,Q^2)\sim
x^{-\Delta(Q^2)}\Bigl({Q^2\over{Q^2+Q_0^2}}\Bigr)^{1+\Delta(Q^2)},
\eeq{eq41} where $\Delta(Q^2)=\alpha_t(0)-1$ may be a constant, in
particular.

Note that since the Regge asymptotics of the Veneziano model is
$\sim(-\alpha' s)^{\alpha(t)-1},$ the only way to incorporate
there the $Q^2$-dependence is through the slope $\alpha'$
\cite{JMP,FJM}, i.e. by making the trajectories $Q^2-$-dependent,
thus violating Regge factorization. The $Q^2$-dependent intercepts
were used earlier \cite{BGP,K} in a different context, namely, to
cope with the observed "hardening" of the small-$x$ physics with
increasing $Q^2$ (Bjorken scaling violation). Although we do not
exclude this possibility (treating it as ``effective'' Regge pole),
we study here the different option of introducing  scaling violation
in the constant $g$ appearing, besides $\alpha'$, in the residue
of DAMA, eq (11).

Using the explicit Regge asymptotic form of DAMA, (\ref{eq35}), and
neglecting the logarithmic dependence of $g$,  we make the
following identification \beq
g(Q^2)^{\alpha_t(0)+a}=\left(Q_{lim}^2
\over{Q^2+Q_0^2}\right)^{\alpha_t(0)}. \eeq{eq45}
Note that eq. (\ref{eq45}) is transcendent with
respect to $g$, since $a=a(g)= Re\
\alpha\Bigl({\alpha_t(0)\over{\alpha'(0)\ln g}}\Bigr)$. Another
point to be mentioned is that this equation is not valid in the whole
range of $Q^2$,
since for $Q^2$ close to $Q_{lim}^2$, $g$ may get smaller than $1$,
which is unacceptable in DAMA. For large $Q^2$, the
$Q^2$-dependence of $\log g$ and $b=b(Q^2)=Im\
\alpha\Bigl({\alpha_t(0) \over{\alpha'(0)\ln g}}\Bigr)$ in eq.
(\ref{eq35}) cannot be neglected, it might contribute to scaling
violation.

\begin{figure}[htb]
        \insertplots{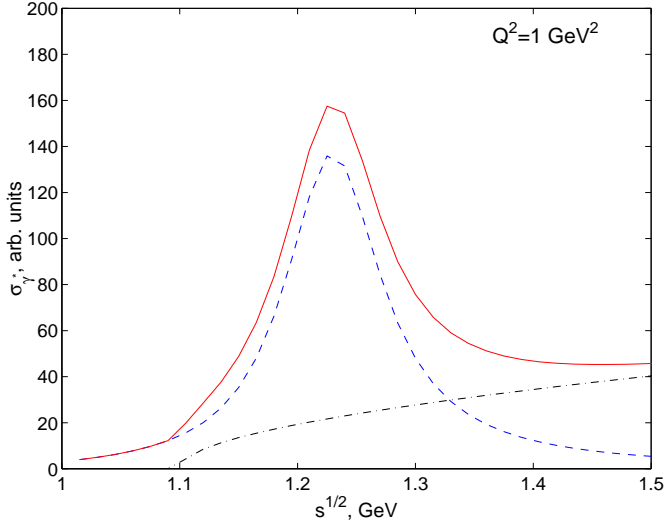}
\caption{ $\gamma^* p$ total cross section as a
function of $\sqrt{s}$. The dashed line shows the contribution
from the $\Delta$ resonance, the dot-dashed line corresponds to
the background, i.e. the contribution from the exotic trajectory.
Here $Q^2=1\ GeV^2$. }
\label{fig2}
\end{figure}

\section{Scaling at large $x$}

Let us now consider the extreme case of a single resonance
contribution.

A resonance pole in DAMA contributes with
$$A(s,t)=g^{n+\alpha_t(0)} {C_n\over{n-\alpha(s)}}.$$
At the resonance $s=s_{Res}$ one has $Re\ \alpha(s_{Res})=n$ and
${Q^2(1-x)\over x}=s_{Res}-m^2$, hence
$$
F_2(x,Q^2)={Q^2(1-x)\over{4\pi^2\alpha\Bigl(1+{4m^2x^2\over
{Q^2}}\Bigr)}} {C_n\over {Im\ \alpha(s_{Res})}}g(Q^2)^{n+\alpha_t(0)}.
$$
As $x\rightarrow 1$
$Q^2\approx{s_{Res}-m^2\over{1-x}}\rightarrow
\infty$ and
$$F_2(x,Q^2)\sim g\Bigl({s_{Res}-m^2\over{1-x}}\Bigr)^{n+\alpha_t(0)}.$$
By using the approximate solution
$
g(Q^2)\approx\left({Q^2_{lim}/
Q^2}\right)^{\alpha_t(0)\over{\alpha_t(0)+a}},
$
where $a$ is a slowly varying function of $g$, we get for $x$ near $1$
$$F_2(x,Q^2)\sim (1-x)^{\alpha_t(0)(n+\alpha_t(0))\over{\alpha_t(0)+a}},
$$
where the limits for $x$ are defined by
$Q_0^2\ll{s_{Res}-m^2\over{1-x}}\leq Q^2_{lim}$.

We recognize a typical large-$x$ scaling behaviour $(1-x)^N$ with
the power $N$ (counting the quarks in the reaction)  depending
basically on the intercept of the $t$-channel trajectories.

\begin{figure}[htb]
        \insertplots{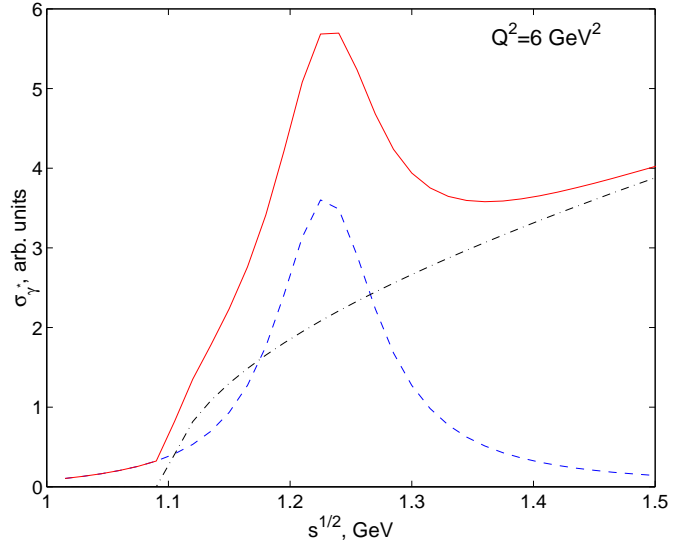}
\caption{ $\gamma^* p$ total cross section as a
function of $\sqrt{s}$. The dashed line shows the contribution
from the $\Delta$ resonance, the dot-dashed line corresponds to
the background, i.e. the contribution from the exotic trajectory.
Here $Q^2=6\ GeV^2$. }
\label{fig3}
\end{figure}

\section{Numerical Estimates}

Having fixed the $Q^2$-dependence of the dual model  by matching
its Regge asymptotic behaviour to that of the structure
functions, we now use this dual model to extrapolate it down to the
resonance region, where its pole expansion (\ref{eq23}) is
appropriate - now complemented with the $Q^2$-dependence through
$g(Q^2)$, fixed by eq. (\ref{eq45})

\begin{table}[htb]
\caption{Parameters used in the calculations shown in
Figs.  \ref{fig1},~\ref{fig2} and \ref{fig3}.}
\label{tab1}

\begin{tabular}{lll}
\hline \noalign{\smallskip}
\hfill \vline&$\Delta$ Resonance \hfill \vline & Background \\
\noalign{\smallskip}\hline\noalign{\smallskip}
$Q_{lim}^2,\ GeV^2$\hfill \vline& 62\hfill \vline  & 120\\
$Q_0^2,\ GeV^2$\hfill \vline&0.01\hfill \vline  & 2.5  \\
\noalign{\smallskip}\hline\noalign{\smallskip}
Dual \hfill \vline&$\alpha_f(t)$ is\hfill \vline &$\alpha_P(t)$ is\\
trajectory \hfill \vline&dual to $\alpha_{\Delta}(s)$\hfill \vline
&dual to $\alpha_E(s)$\\
\noalign{\smallskip}\hline\noalign{\smallskip}
\hfill \vline &$\alpha_f(0)=0.9$\hfill \vline & $\alpha_P(0)=1+0.077\cdot$\\
\hfill \vline &\hfill \vline  &$\cdot\left(1+\frac{2Q^2}
{Q^2+1.117}\right)$ \cite{K}\\
\noalign{\smallskip}\hline
\end{tabular}\\

Normalization coefficient $c=0.03$.
\end{table}

As has been already said, we write the imaginary part of the scattering
amplitude as the sum of two terms - a diffractive (background) and
non-diffractive (resonance) one. Note that $g(Q^2)$ has the same
functional form (\ref{eq45}) in both cases, only the values of the
parameters differ (they are fixed from the small-$x$ fits
\cite{JMP99} of the SF).

\begin{figure*}[htb]
        \insertplot{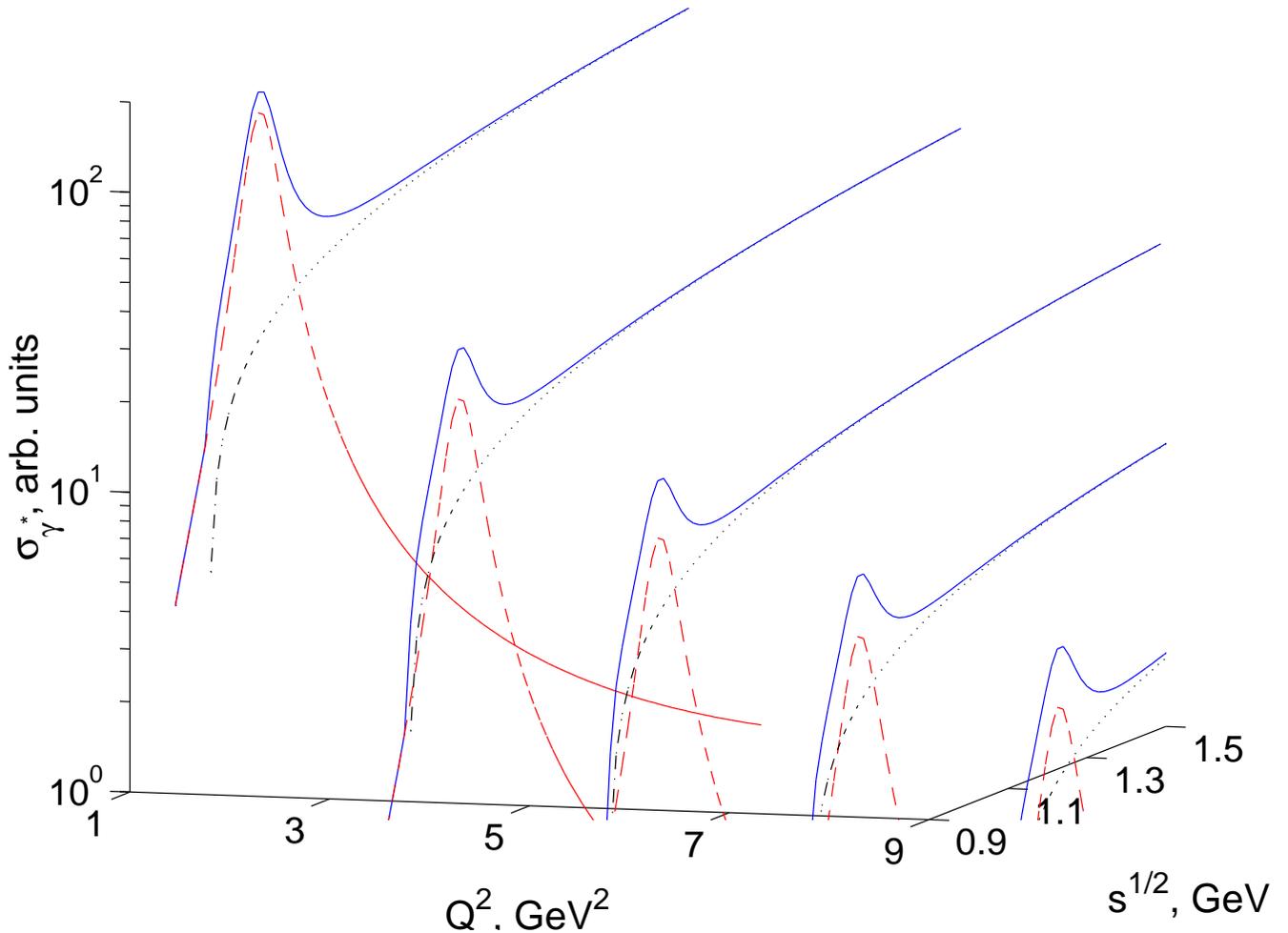}
\caption{ $\gamma^* p$ total cross section as a
function of $\sqrt{s}$ and $Q^2$. For different values of $Q^2$
we show the contributions
from the $\Delta$ resonance (dashed line), the background, i.e.
the contribution from the
exotic trajectory (dot-dashed line)
and their sum (full line).} \label{fig4}
\end{figure*}

At low, resonance, energies the $\gamma^* p$ scattering exhibits a rich
resonance structure intensively studied in a number of papers.
About 20 resonances overlap, their relative importance varying
with $Q^2$, but only a few can be identified more or less
unambiguously. These are: $\Delta^+(1236)$ with $J^P={3^+\over 2}$,
$N^{*+}(1520)$, $J^P={3^-\over 2}$, $N^{*+}(1688)$,
$J^P={5^+\over 2}$ and $N^{*+}(1920)$ with $J^P={7^+\over 2}.$ They
lie on the $\Delta$ and the exchange-degenerate $N$ trajectories.
In this work we are mainly interested in introducing
$Q^2$-dependence into the scattering amplitude, therefore we 
concentrate on a single resonance ($\Delta^+(1236)$) at different
values of $Q^2$. We use trajectories (\ref{eq33}) in which the
lowest pion-nucleon threshold is included explicitly, while higher
thresholds are approximated by a linear term:
$$\alpha_{\Delta}(s)=0.1+0.84s+0.1331(\sqrt{s_0}-\sqrt{s_0-s}),$$
where $s_0=(m_\pi^2+m_N^2)$.\footnote{ Actually, trajectories
without any linear term (see e.g. \cite{FJMPP}) could be more
appropriate (and will be studied in future).} The above values of
the parameters are chosen so as to fit the known mass and width of the
$\Delta$ resonance in a way consistent with the known linear
parameterizations.

In the interval of interest $\sqrt{s}=1.1-1.5\ GeV,\ t=0$, we have
$u=m_N^2-s<0$, so it is far from resonance region, therefore we
neglect the contribution from $D(u,t)$ for both the resonance and
the background terms.

The smooth background is also modeled by a single term and exotic
trajectory (\ref{eq32}). As has been already explained, the direct channel 
Regge pole does not produce here physical resonances. The parameters of the 
exotic trajectory are: \beq 
\alpha_E(s)=-0.25+0.25(\sqrt{1.21}-\sqrt{1.21-s}), \eeq{ex} where
$s_E=1.1^2\ GeV^2$ is an effective exotic threshold. Obviously "pole"
does not mean a resonance in this case.

Figure \ref{fig1}
shows $g$ as a function of $Q^2$ for $\Delta$ and exotic
trajectories.
The resulting cross sections (imaginary part of the amplitude) in
the resonance region is shown in Figs. \ref{fig2} and \ref{fig3}
for two values of $Q^2=1$ and $6\ GeV^2$. It is in qualitative
agreement with the experimental data \cite{JM}.
Figure \ref{fig4} shows the dual
properties of the cross section in two dimensions - one is the
energy squared $s$ and the other one is virtuality $Q^2$.
Table \ref{tab1} shows the values
of the parameters used in our calculations.

The main conclusions from our analysis are as follows:\\
A) the $Q^2$-dependence at low- and high-x (or high- and low-s) are
interrelated and of the same origin;\\
B) even a single (low energy) resonance can produce a smooth
scaling-like curve in the structure function (parton-hadron
duality).

To summarize, we have suggested an explicit dual model in which
the $Q^2$-dependence introduced in the low-$x$  domain is extended
to the whole kinematic region, in particular, to the region of resonances.
The resulting predictions for the first resonance in the
$\gamma^* p$ system shown in Figs. \ref{fig2},~\ref{fig3} are in
quantitative agreement with data.

\section{Acknowledgments}
L.J. thanks INFN and the Torino University, where this work was completed,
for their hospitality and support.  L.J. ans V.M. acknowledge the support
by INTAS, Grant 00-00366.
The work of L.J. was supported also by the US Civilian Research and
Development Foundation (CRDF), Grant UP1-2119.


\begin{thebibliography}{}
%
%
\bibitem{Niculescu} I. Niculescu et al., Phys. Rev. Letters \textbf{85}, (2000)
1182, 1186.

\bibitem{JMP} L.L. Jenkovszky, V.K. Magas and F. Paccanoni,
Proceedings of the "New Trend in High-Energy Physics", Crimea,
Ukraine, May 27 - June 4, 2000, p. 121.

\bibitem{FJM} R. Fiore, L. Jenkovszky, V. Magas, Proceedings of the
"Diffractin-2000", Cetraro, Italy, 2-8 September 2000, Nucl. Phys.
B (Proceedings Supplements) \textbf{99}, (2001) 131.

\bibitem{JM} L. Jenkovszky and V. Magas, to be
published in the Proceedings of the 9th Blois Workshop On Elastic and Diffractive
Scattering, June 9-15, 2001, Prague, Czech Republic;
to be published in the Proceedings of "Spin-2001", 9th International Workshop On High-Energy Spin Physics,
2-7 Aug 2001, Dubna, Russia; to be
published in the Proceedings of the "ISMD 2001",
XXXI International Symposium on Multiparticle
Dynamics, September 1-7, 2001,  Datong, China, hep-ph/0111398;
L. Jenkovszky, T. Korzhinskaya, V. Kuvshinov and V. Magas, to be
published in the Proceedings of the "New Trend in High-Energy Physics",
Yalta, Crimea, Ukraine, Sept 22-29, 2001; L. Jenkovszky,
V.K. Magas, E. Predazzi, to be published in the
Proceedings of the 6th International Summer
School-Seminar On Actual Problems Of High Energy Physics,
Gomel, Belarus, August 7-16, 2001, nucl-th/0110085.

\bibitem{Veneziano}G. Veneziano, Nuovo Cim. \textbf{A 57}, (1968) 190.

\bibitem{BG}  E.D. Bloom, E.J. Gilman, Phys. Rev. Letters \textbf{25},
(1970) 1149; Phys. Rev. \textbf{D 4}, (1971) 2901.

\bibitem{FH} P. Freund, Phys. Rev. Letters \textbf{20}, (1968) 235; H. Harari,
Phys. Rev. Letters \textbf{20}, (1968) 1395.

\bibitem{RR} V.~Rittenberg and H.R.~Rubinstein, Nucl.
Phys. \textbf{B 28}, (1971) 184;

\bibitem{Schierholz} G.~Schierholz and M.G.~Schmidt, Phys. Rev.
\textbf{D 10}, (1974) 175.

\bibitem{carlson} A. De Rujula, H. Georgi, and H.D. Politzer,
Ann. Phys. (N.Y.) \textbf{103}, (1977) 315;  P. Stoler, Phys. Rev.
Letter \textbf{ 66}, (1991) 1003; P. Stoler, Phys. Rev. \textbf{ D 44},
(1991) 73; I. Afanasiev, C.E. Carlson, Ch. Wahlqvist, Phys. Rev. \textbf{D 62},
(2000) 074011; F.E. Close and N. Isgur, Phys. Letter \textbf{B 509}, (2001) 81;
N. Isgur, S. Jeschonnek, W. Melnitchouk, and J.W. Van Orden,
Phys. Rev. \textbf{D 64}, (2001) 054004.

\bibitem{Carl}C.E.~Carlson, N. ~Mukhopadhyay, Phys. Rev. \textbf{D 41}, (1990)
2343.

\bibitem{DHS} A.A. Logunov, L.D. Soloviov, A.N. Tavkhelidze Phys. Lett. \textbf{ B 24}, (1967)
181; R. Dolen, D. Horn and C. Schmid  Phys. Rev. \textbf{ 166}, (1968)
1768.


\bibitem{DG} Marc Damashek and Frederick J.~Gilman, Phys. Rev.
\textbf{ D 1}, (1970) 1319.

\bibitem{Nachtmann} O. Nachtmann, Nucl. Phys. \textbf{B 63}, (1973) 237;
\textbf{B 78}, (1974) 455.

\bibitem{BEG} A. Bramon, E. Etim and M. Greco, Phys. Letters \textbf{B 41}, (1972) 609.

\bibitem{EM} E. Etim, A. Malecki, Nuovo Cim \textbf{ A 104}, (1991) 531.


\bibitem{D} A. Bugrij et al., Fortschr. Phys., \textbf{ 21},
(1973) 427.

\bibitem{DP} A. Degasperis and E. Predazzi, Nuovo Cim. \textbf{ A 65}, (1970)
764.

\bibitem{FJMPP} R. Fiore, L. Jenkovszky, V. Magas, F. Paccanoni
and A. Papa, hep-ph/0011035, Eur. Phys. J. \textbf{ A 10}, (2001) 217.

\bibitem{BGP} M. Bertini, M. Giffon and E. Predazzi, Phys. Letters
\textbf{ B 349}, (1995) 561.

\bibitem{K} A. Capella, A. Kaidalov, C. Merino, J. Tran Thanh Van,
Phys. Letters \textbf{B 337}, (1994) 358; L.P.A. Haakman, A. Kaidalov,
J.H. Koch, Phys. Letters \textbf{ 365}, (1996) 411.

\bibitem{JMP99} L. Jenkovszky, E. Martynov, F. Paccanoni, Padova
preprint PFDPD 95/TH/21; P. Desgrolard, A.I. Lengyel, E.S.
Martynov, EPJ \textbf{C 7}, (1999) 655.




\end{thebibliography}
%

\end{document}